\documentclass[useAMS,usenatbib]{mn2e}
\usepackage{ upgreek } 
\usepackage{graphicx}
\usepackage{caption}
\usepackage{subcaption}
\usepackage{amssymb}
\usepackage{amsmath}
\usepackage{float}
\usepackage{color, array}
\usepackage{booktabs} 
\usepackage{lscape}

\title[Possibility of Primordial black holes  Collision with Earth and the Consequences]{Possibility of Primordial black holes  Collision with Earth and the Consequences of this Collision}

\author[S. Rahvar]{Sohrab Rahvar\thanks{E-mail: rahvar@sharif.edu} \\
$^{1}$Physics Department, Sharif University, P.O.Box 11365-9161, Azadi Avenue, Tehran, Iran}
\pagerange{\pageref{firstpage}--\pageref{lastpage}} \pubyear{2021}

\def\LaTeX{L\kern-.36em\raise.3ex\hbox{a}\kern-.15em
T\kern-.1667em\lower.7ex\hbox{E}\kern-.125emX}

\begin{document}
\label{firstpage}

\maketitle 

\begin{abstract}
In the context of the existence of Primordial black holes  (PBHs), they may compose a fraction of the dark matter of the Universe.  Assuming that PBHs fill the dark content of the Milky Way Galaxy in the  Galactic halo and dark disk, we calculate the probability of collision of the PBHs with Earth. This collision has different consequences as heating the interior of the earth through dynamical friction and accretion processes. In this work, we calculate the rate of collisions and a fraction of black holes  that can be trapped inside the earth.
Finally, we compare the danger of PBH collision with the asteroid impacts on Earth.
      
\end{abstract}

\section{Introduction}
The existence of black holes  as the singularities of space-time is one of the predictions of general relativity \citep{pen}. These objects have been observed indirectly for decades by different methods \citep{samimi,ship}. Recent observations of gravitational waves by the LIGO and Virgo detectors revealed the existence of massive binary black holes  where merging of black holes  generate the burst of gravitational waves \citep{ligo1,ligo2}. The mass of binary black holes  derived from the analysis of gravitational wave data indicates that the mass should be around tens of solar masses \citep{gwmass2,gwmass1}. It has been known that the mass of black holes  from the astrophysical process is around a few solar masses and this mass is much small compared to the mass of binary black holes derived from the LIGO observations. One of the possibilities for the production of these large mass black holes  could be through the process of Primordial Black Hole (PBH) formation in the early universe \citep{p1}. The interpretation of LIGO data based on this hypothesis is under investigation \citep{rah}.

There are several scenarios to describe the formation of PBHs as the collapse of 
over dense regions in the early universe where in this case we expect a wide mass function for 
PBHs \citep{mf1,mf2,carr}. 
Parts of window for the mass function of the PBHs have been excluded by various observations such as microlensing observations of compact objects along the line of sight of Large and Small Magellanic Clouds and Galactic Bulge \citep{anna}. However, there are still two open regions for this window for the small mass (i.e. $M< 10^{-7} M_\odot$) and large masses (i.e. $M>10^5 M_\odot$).  While the black holes are evaporating by 
the Hawking radiation, however large mass black holes from the early times can still survive from the radiation. Assuming the existence of PBHs from the early Universe, in this work, we examine the scenario of 
collision of the PBHs with the earth and investigate the consequences of this collision. 

In Section (\ref{Srate}),  we calculate the rate of PBHs collision with the Earth as a function of their masses. In Section (\ref{heating}), we investigate the dissipation processes from the collision of back holes  with earth. Also we calculate the probability of trapping black holes at the interior of the earth. In Section (\ref{energy}) we study the energy release during the collision of black holes  with earth and compare it with the impact of asteroids.  The conclusion is given in (\ref{conc}).



\section{Rate of PBH collision with Earth}
\label{Srate}
Let us assume a fraction of dark matter is made of PBHs with a mass function of $f(M)$ for PBHs. For simplicity in the calculation, we take a Dirac-Delta mass function  $\delta(M-M_{0})$  for the black holes, however, we will consider a mass function in this calculation when we are calculating the overall rate of collisions with earth. For the mass density of dark matter, $\rho_{DM}(x)$, the number density of black holes  is $n_{bh} = f\rho_{DM}(x)/M$ where $f$ is the fraction of halo is made of black holes . For PBHs moving inside the dark halo with the dispersion velocity associated with the objects inside the halo, PBHs may collide with the stars, planets, and the earth. Since the dark component of our galaxy is in the halo and dark disk, the collisions can happen from these two different structures. 

We take a black hole as a projectile approaching the earth. The maximum amount of angular momentum of a black hole at the closest approach to earth is $\ell = R_e  mv_{max}$ where $v_{max}$ is the maximum velocity that the black hole can achieve in a tangent collision with earth and $R_e$ is the radius of the earth. On the other hand, the angular momentum of the black hole with respect to earth at infinity is $\ell = b m v$ where $b$ is the impact parameter and $v$ is the relative velocity of the black hole with respect to the earth at infinity. Using the conservation of angular momentum and conservation of energy, the 
impact parameter obtain as 
\begin{equation}
b = R_e (1 + \frac{2GM_e}{R_ev^2})^{1/2},
\label{impact}
\end{equation}
where $M_e$ is the mass of Earth. The cross-section associated to this collision is $\sigma = \pi R_e^2( 1 + {v_{es}^2}/{v^2})$ where $v_{es}$ is the escape velocity from earth. From the cross-section, the number density of black holes in space, and relative velocity of black holes with respect to the earth, we can calculate the impact rate of black holes with the earth. Let us take a normalized probability distribution function for the velocity of black holes  as $p(v)$,  the rate of collisions of the black holes  with the earth is  
\begin{equation}
\frac{dN}{dt} = \int J(v) \sigma(v) p(v) dv^3,
\label{dn}
\end{equation}
where $J(v) = n_{bh} v$ is the current of black holes  and 
\begin{equation}
p(v) = \frac{1}{(2\pi \sigma^2)^{3/2}}\exp(-v^2/2\sigma^2)
\end{equation}
 is the normalized distribution function of objects in the Galactic halo \citep{gd}. The result of integration is 
\begin{equation}
\frac{dN}{dt} = \sqrt{2\pi} n_{bh} R_e^2 \frac{(v_{es}^2 + 2\sigma^2)}{\sigma}.
\label{dndt}
\end{equation}

The local dark matter of Galaxy is composed from the galactic halo structure and local dark disk. The standard halo model provides the density of halo dark matter to be $\rho_h = 8\times 10^{-3} M_\odot pc^{-3}$ with the dispersion velocity of objects inside halo, $\sigma = 220~ \text{km s}^{-1}$ \citep{gd}.  On the other hand  the investigation on the dynamics of stars perpendicular to the galactic disk with the Gaia data provides 
the local density of Dark Disk (DD), $\rho_{DD} = 0.016 M_\odot pc^{-3}$ \citep{dd}. The dispersion velocity of thin disk model which is compatible with the other observations are as $\sigma_r = 27.4~ \text{km s}^{-1}$, $\sigma_\theta = 20.8~ \text{km s}^{-1}$ and $\sigma_z = 16.3~\text{km s}^{-1}$ where overall dispersion 
$\sigma^2 = \sigma_r^2 + \sigma_\theta^2 +  \sigma_z^2$ is $\sigma = 38.1~\text{km s}^{-1}$ \citep{moniez}.  We adapt this dispersion velocity for the black holes  in the dark disk.   Substituting the numerical values in equation (\ref{dndt}), the rate of collision of the black holes  with earth from the galactic halo and dark disk is  
\begin{eqnarray}
\label{rate1}
(\frac{dN}{dt})_{halo} &=& 7.2\times 10^{-9} f (\frac{M}{10^{23} gr})^{-1} Gyr^{-1}, \\
(\frac{dN}{dt})_{DD} &=& 2.6\times 10^{-9} f (\frac{M}{10^{23} gr})^{-1} Gyr^{-1},
\label{rate2}
\end{eqnarray}
where adding the contribution of the halo and dark disk, the overall rate of collision is
\begin{equation}
\frac{dN}{dt} \simeq 10^{-8} f  (\frac{M}{10^{23} gr})^{-1} Gyr^{-1}.
\label{rate}
\end{equation}
In order to have at least one collision of PBHs with earth per one billion year, we have to put the constrain 
of $M< 10^{15} f $~gr (with the corresponding Schwarzschild radius of $R_s \lesssim 10^{-12}$ mm). The observational constrain for PBHS in this mass range is discussed in \citet{carr2020} where it is possible to allow regions of the parameter space with $f<1$.

The black holes  lose their masses through the Hawking radiation as  \citep{h1}
\begin{equation}
\frac{dM}{dt} =-  \frac{\hbar c^4}{1530 \pi G^2} \frac{1}{M^2}
\end{equation}
where the solution of this equation results in $M(t) = (M_i^3 - Ct)^{1/3}$  where $C = 10^{32}$ gr/yr and minimum initial mass of black holes  that survive at present time is  $M_{i}>10^{14}$ gr. 
Let us take a lognormal initial mass function of PBHs as \citep{log1,log2,log3}
\begin{equation}
\psi(M_i) = \frac{1}{\sqrt{2\pi}\sigma M_i} \exp[-\frac{\log^2(M_i/M_c)}{2\sigma^2}],
\end{equation}
where $M_c$ represents the peak of $\psi(M_i) M_i$ function. Black holes  with the mass in the order of $10^{14}$ gr will lose half of their masses during the life-time of the Universe. So for the initial black holes within the range of $10^{14} \text{gr}<M_i<10^{15}$gr we would expect black holes  survived from the Hawking evaporation and have one collision per $10^9$ year.



 We can imagine various physical processes for the collision of the low-mass PBHs with earth. The effect of this collision on earth may have geological traces as well as effects on living species on earth. During the passage of PBH from the earth (i) black holes  can decelerate due to dynamical friction, also (ii) accrete matter from the interior of the earth on it. 
The deceleration due to dynamical friction is a consequence of gravitational interaction with the material of earth as well as the momentum transfer due to the accretion of matter on it. On the other hand, we expect the radiation due to the accretion where the overall effects could generate heat at the interior of the earth.

 We define a deceleration time scale ($t_d$) during the passage of a PBH through the earth. For the case that it is shorter than the crossing time scale $t_c$ of the PBH through the earth (i.e. $t_d<t_c$) the PBH will lose its energy, spiral, and sink into the center of the earth and finally could be trapped by the earth. The accretion of matter on the black hole will heat the interior of the earth and makes the black hole grow and finally, all the earth will be swallowed by the black hole.  We will discuss the threshold mass of black holes  that can be trapped inside the earth and calculate the rate of these events compared to the lifetime of the Universe. For the case that PBH crosses the earth and leave from the other side to space, black holes  just transfer part of its energy to the earth. 
      
 
 \begin{figure*}
 	\begin{center}
 				\includegraphics[width=7.5cm ]{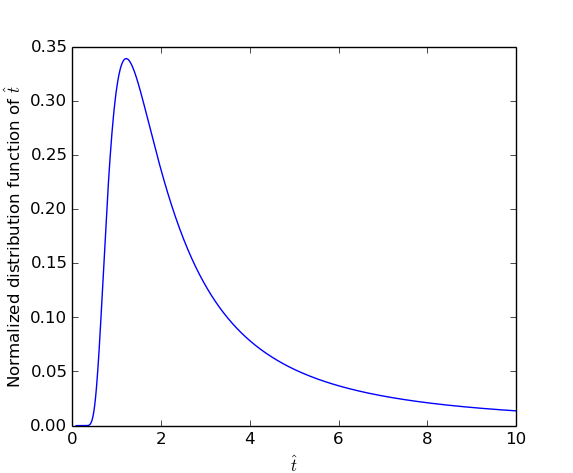}
 		\caption{Normalized distribution function for the normalized crossing-time of black holes (i.e. $\hat{t} = t/t_c)$ that cross the earth during collision.}
		\label{fig1}
 	\end{center}
 \end{figure*}
 
  We define the crossing time scale of black holes  through the earth as $t_c = R_e/{\bar v}$ where the average velocity in terms of dispersion velocity from the Maxwell distribution is ${\bar v} = \sqrt{3} \sigma$. For the dispersion velocity of halo $\sigma (halo) = 220~ \text{km~s}^{-1}$ and dispersion velocity of dark disk $\sigma (DD) = 38~ \text{km~s}^{-1}$, the corresponding crossing time scales are $t_c(halo) = 17$ s and $t_c(DD) = 97$ s, respectively. We can calculate the distribution function for the normalized transit time by defining $\hat{t} = (R_e/v)/t_c = \sqrt{3} \sigma/v$. We use the Maxwell-Boltzmann distribution for the velocity of black holes, then the distribution function for $\hat{t}$ obtain as 
\begin{equation}
p(\hat{t}) d\hat{t} = \sqrt{\frac{6}{\pi}} \frac{1}{\hat{t}^2}\exp{(-\frac{3}{2\hat{t}^2})} d\hat{t}
\end{equation}
Figure (\ref{fig1}) shows this distribution function in terms of normalized transit time-scale. In the next section, we will compare the transit time-scale with the time-scales of dissipation processes to evaluate the impact of a black hole  on the Earth.


 \section{PBH crossing and dissipation processes}
 \label{heating}

The dissipation process while PBH crossing the earth happens due to two different mechanisms of dynamical friction and accretion of the matter by a black hole. The dynamical friction is a result of the conservation of energy and momentum of a massive projectile and the collection of particles in the medium while the projectile moving through it \citep{chan}. The result is the deceleration of the massive projectile and increasing the kinetic energy of the medium.

The deceleration due to the dynamical friction as a result of the interaction of matter in the medium by a black hole is given by 
\begin{equation}
\frac{d\mathbf{v}_{bh}}{dt} = -\frac{4\pi \ln (\Lambda) G^2  \rho_e M}{v_{bh}^3}\left[\mathrm{erf}(X)-\frac{2X}{\sqrt{\pi}}e^{-X^2}\right]\mathbf{v}_{bh},
\label{dfric}
\end{equation}
 where $ X = v_{bh}/(\sqrt{2} \sigma_e)$ and $\sigma_e$ is the dispersion velocity of particles inside the earth, $\rho_e$ is the density of earth and $\ln\Lambda$ is the column logarithm and is given by 
 \begin{equation}
 \ln\Lambda = \ln\frac{b_{max} \sigma_e^2}{G(M + m)}, 
 \label{lam}
 \end{equation}
 where $b_{max}$ is the maximum impact parameter that we adapt the radius of earth (i.e.  $b_{max} = R_e$) and $m$ is the mass particles inside the earth. Equation (\ref{lam}) simplifies to $\ln\Lambda = \ln(R_e\sigma_e^2/(GM))$ where substituting the numerical values results in 
 \begin{equation}  
  \ln\Lambda = 10.3 + \ln(\frac{M}{10^{23}\text{gr}}).
  \end{equation}
  The time-scale for deceleration of black holes  while passing through the earth as a result of dynamical 
 friction is defined by $t_{DF} = v_{bh}/\dot{v}_{bh}$. By taking into account that $v_{bh}\gg \sigma_e$, the 
 time scale simplifies to 
 \begin{equation}
 t_{DF} = \frac{v_{bh}^3}{4\pi G^2 M \rho_e \ln\Lambda}, 
 \end{equation}
 where for the black holes from the halo and dark disk, the numerical values of this time scale is  
 \begin{equation}
 t_{DF} = C (\frac{M}{10^{23}~\text{gr}})^{-1} (1 + \frac{1}{10}\ln\frac{M}{10^{23}~\text{gr}})^{-1} s,
 \end{equation}
 where the parameter $C$ depending on the black holes from the halo or dark disk is, $C_{halo} = 3.8\times 10^{10}$ and $C_{DD} = 1.95\times 10^{8}$, respectively. 
 
 Here we compare the dynamical friction time-scale ($t_{DF}$) with the crossing-time scale of the black holes through the earth ($t_c$). Let us define the parameter of $Y = t_{DF}/t_c$ where substituting the parameters results in 
 \begin{equation}
 Y = \frac{v_{bh}^4}{A^4}.
 \end{equation}
 where $A^4 = 4\pi R_e G^2 M \rho_b \ln\Lambda$. 
 We have two phases for $Y$ where for $Y>1$ dissipation is not effective and black holes can cross the earth without sinking inside the earth and for $Y<1$ dissipation is effective. For the velocity of black holes  in the halo ($dN/dv_{bh}$) follows a Maxwell-Boltzmann distribution,
 \begin{equation}
 \frac{dN}{dv_{bh}} dv_{bh}= \frac{4\pi v_{bh}^2}{(2\pi\sigma^2)^{3/2}}  \exp(-\frac{v_{bh}^2}{2\sigma^2}) dv_{bh},
 \end{equation}
 we change the variable from $v_{bh}$ to $Y$, the result is  
 \begin{equation}
 \frac{dN}{dY} dY= \frac{1}{\sqrt{2\pi}}\kappa^3 Y^{-1/4} \exp(-\frac{\kappa^2 Y^{1/2}}{2}) dY,
 \end{equation}
 where  $$\kappa = A/\sigma = 6.25\times 10^{-3} (\frac{\sigma}{220 \text{km/s}})^{-1} (\frac{M}{10^{23}\text{gr}})^{1/4}.$$
 
Integrating over $Y$ from zero to one, results in the probability of black holes  sink in the earth.  
The numerical value of this probability is 
$$P(Y<1) = 1.3\times 10^{-7} (\frac{\sigma}{220 \text{km/s}})^{-3} (\frac{M}{10^{23}\text{gr}})^{3/4}.$$
If we multiply the probability function to the total number of collisions of black holes  originated from the disk and halo (equations (\ref{rate1}) and (\ref{rate2})), we obtain the rate of black hole collisions with earth that can be sinked in the earth  
\begin{eqnarray}
\label{nrate1}
(\frac{dN}{dt})_{h}(Y<1) &=& 9.4\times 10^{-16} f (\frac{M}{10^{23} gr})^{-1/4} \text{Gyr}^{-1}, \\
(\frac{dN}{dt})_{d}(Y<1) &=& 1.3\times 10^{-13} f (\frac{M}{10^{23} gr})^{-1/4} \text{Gyr}^{-1}.
\label{nrate2}
\end{eqnarray}

Since the rate of black holes that could be trapped in the earth, originated from the halo is three orders of magnitude smaller than that of the dark disk, we can ignore the contribution from the halo. For black holes  with the mass of $M\sim 10^{15}$gr, assuming that $f =1$, we expect the rate of black holes  that can be trapped in the earth is $(dN/dt)(Y<1) \simeq 10^{-11} \text{Gyr}^{-1}$. Taking into account the possible smallest black holes with the Planck mass $M = 10^{5}$ gr, where charged black holes can survive from the Hawking evaporation \citep{bal}, the rate of smallest black holes that can be trapped inside the Earth during the lifetime of the Earth is almost zero. Concluding this section, we ensure that all the black holes colliding with earth will cross its interior and leave space.

\section{The black holes  interaction and energy release}
 \label{energy}
In the previous section, we have shown that the probability that a black hole collision with the earth and can be trapped at the interior of the Earth is almost zero.  However, those black holes that cross the earth can heat the interior of the earth with various mechanisms. 

 In what follows we investigate the amount of energy release of black holes in addition to the amount of accretion during the passage of the black holes through the earth.  The lose of kinetic energy during this passage (from Newton's second law) which transfers to heat is equal to the mass of PBH times the deceleration times the length that PBH crossing the earth (i.e. $Q \simeq M\dot{v}_{bh} R_e$). Using equation (\ref{dfric}) for the deceleration of black holes  due to dynamical friction and using the numerical values, the average heat production is  
 \begin{eqnarray}
 {\bar Q _{Halo}}&=& 3.7\times 10^{21} (\frac{v_{bh}}{220 \text{km~s}^{-1}})^{-2}(\frac{M}{10^{23}~\text{gr}})^2~ J, \nonumber \\
 {\bar Q_{disk}} &=& 1.2\times 10^{24} (\frac{v_{bh}}{38 \text{km~s}^{-1}})^{-2}(\frac{M}{10^{23}~\text{gr}})^2~J,
 \label{df}
 \end{eqnarray}
where depending on the velocity of black holes from the halo or dark disk the energy loss would be different. For the mass range of $M<10^{15} \text{gr}$, the number of interactions of primordial black holes with the earth is larger than one event per Gyr where for this range of masses the heat production due to the dynamical friction is $Q_{halo}<10^5~J$ and $Q_{disk}<10^8~J$. 

The other source of heat production is the accretion of matter into the black holes while passing through the Earth. The maximum amount of energy released from the accreting matter around the black hole is given by the Eddington limit \citep{edd}. There are different sub-Eddington accretion
mechanisms that lead to lower luminosities by several orders of magnitude \citep{shak}.

For the  Eddington limit, equating the radiation pressure with the gravitation pressure results in 
\begin{equation}
L_{Edd} = \frac{4 \pi G M m_p}{\sigma_{Th}},
\end{equation}  
where $\sigma_{Th}$ is Thompson cross section for photons and electrons. Using 
the numerical values, the Eddington limit is  
\begin{equation}
L_{Edd} = 6.5\times 10^{27} (\frac{M}{10^{23}\text{gr}})~ \text{erg/s},
\end{equation}
For a black hole with the mass in the order of $M = 10^{15} \text{gr}$, using the crossing time-scale 
 through the earth (i.e. $t_c = R_e/\bar{v}$), the corresponding energy from the radiation, $Q = L_E t_c$ 
is 
\begin{eqnarray}
\label{ed1}
Q_{Halo}^{(rad)} &=& 2.2\times 10^{22}  (\frac{M}{10^{23}\text{gr}}) J,  \\
Q_{disk}^{(rad)} &=& 1.3\times 10^{23}  (\frac{M}{10^{23}\text{gr}}) J, 
\label{ed2}
\end{eqnarray}
where for the mass range of $M<10^{15}$ gr, the heat production due to accretion in the black holes sourced from halo and disk would be  $Q_{halo}<10^{14}$ J and $Q_{disk}<10^{15}$ J.  Comparing the energy release from the accretion with the dissipation of energy due to dynamical friction in equation (\ref{df}), we can neglect the heat production due to the dynamical friction.

In order to have an idea about the amount of energy that is released from the black hole impact, we compare it with the energy released from the asteroid that impacts the earth. For an asteroid impact with the crater size of $D$ on earth, the energy release is about $ E \simeq 10^{16} (D/\text{km})^3$ J \citep{Dence}. We take into account $\bar{v}_{astroid} = 20800$ ms$^{-1}$ as the mean impact velocities of asteroids with earth \citep{david}; for a nonelastic impact the energy release would be $E_{astroid} = 2\times 10^{28} (M/10^{23} gr)$J. Comparing the energy release by the asteroid impact with the same mass black hole, asteroids produce five orders of magnitude more energy than that of a black hole.

 On the other hand, the rate of asteroid the impacts, based on estimating the number of impact craters that are presented at Earth's surface is  $\Gamma_{astroid} = 6.64 \times 10^4 (D/{\text{km}})^{2.55} \text{Gyr}^{-1}$ \citep{rate} where for an asteroid with the crater size of km, the energy release is almost equal to the impact of a black hole  with the mass of $M\simeq 10^{15}$ gr. However, we note that the rate of an asteroid impact is four orders of magnitude larger than that of PBH impact. We note that in our estimation, we took $f=1$, however using the observational constrains for PBHs with the $M=10^{15}$gr provides $f\simeq 10^{-5}$ \citep{carr2020} which results in event much higher probability for the asteroid impacts compare to the PBHs. 
The impact of the asteroid and a black hole with the earth also has two different physical features. While an asteroid during the entrance to the atmosphere burn and has an explosive impact on the surface of the earth, we expect that a black hole has different features in the impact.  

For a black hole with the size of Schwarzschild radius, $R_s$ and the relative velocity of $v_{bh}$ with respect to the earth, the cross-section that materials inside the Earth accrete into the horizon of a black hole, similar to equation (\ref{impact}), using the conservation of energy and momentum of the particles that accreting into a black hole is 
\begin{equation}
\label{acc}
\sigma_{acc} = \pi R_s^2 ( 1 +\frac{2GM}{R_sv_{bh}^2}),
\end{equation}
where using $1 +\frac{2GM}{R_sv_{bh}^2}\simeq c^2/v_{bh}^2$ and the numerical values, equation  (\ref{acc}) simplifies to  
\begin{equation}
\sigma_{acc} = 1.6\times 10^{-7} (\frac{M}{10^{23}\text{gr}})^2 \text{m}^2 .
\end{equation}
This accretion of matter results in radiation which is given by equations (\ref{ed1}) and (\ref{ed2}). 

Since the black holes are moving through to the earth, the radiating energy around the Schwarzschild radius transfer away from the trajectory of black holes across a cylinder. This radiation can increase the temperature of the earth in the crust around the cylinder with the center of the trajectory of black holes by the amount of $Q =  m C \Delta T$ where $C$ is the heat capacity of the crust. The majority of crust is made of 
 feldspars \citep{earth1} where we can assign the specific heat capacity of $C \simeq 700$ J kg$^{-1}$ K$^{-1}$ and the melting point of $T_M \sim 1500 K$ \citep{earth2}.

The total mass of the crust that is melted during this heating is $m = Q/(C\Delta T)$, where using the specific heat capacity and the melting temperature for crust, the amount of material that can be melted due to heat production by the accretion from equation (\ref{ed1}) and (\ref{ed2}) is 
\begin{eqnarray}
\label{m1}
m^{(melt)}_{Halo} &=& 1.8\times 10^{16}  (\frac{M}{10^{23}\text{gr}}) \text{~kg},  \\
m^{(melt)}_{disk} &=& 1.2\times 10^{17}  (\frac{M}{10^{23}\text{gr}}) \text{~kg}.
\label{m2}
\end{eqnarray}
From equation (\ref{rate}), for the mass range of $M<10^{15}$ gr, the probability of black holes  collision is larger than one collision per Giga year for $f=1$ where the corresponding melting mass is $m^{(melt)}\simeq  1.2\times 10^9$kg which is approximately equal to the mass of a hill with the hight of $50$ meter. 

  We assume a typical trajectory of the black holes equal to the radius of earth. The density of interior matter is in the order of $\rho = 3$gr/cm$^3$ \citep{earth3} and a cylinder that can be melted due to the passage of black holes has a radius of $R \sim 10$ cm (with the melting mass of $m^{(melt)}\simeq  1.2\times 10^9$kg) which is a result of heating process by the radiation from accretion.  
The geological investigations of the deep drilling program on earth crust show a rapid temperature gradient in the crust \citep{dw}. We expect that such a long melting hole due to the passage of black holes  will rapidly dissipate the energy and earth crust will be recovered where just in the upper parts of curst we may distinguish an unusual defect of the melted rocks along a straight line. The absence of this kind of traces can put constrain on the mass and the abundance of PBHs.
\section{Conclusion}
\label{conc}
In this work, we investigated the scenario of collision of primordial black holes with earth. We calculated the rate of impact assuming that hundred percent of dark matter is made of these objects. Assuming a Dirac-Delta mass function, for the black holes that survived from the Hawking radiation, black holes with the mass $M<10^{15}$ gr have the chance of more than one impact per billion years with Earth. Since the velocity of black holes in the halo and the dark disk is high, the dissipation process inside the earth is not effective to decelerate the black holes and black holes do not sink inside the earth where in this case black holes could heat the interior of the earth and finally swallow the whole mass of earth. The number of this event for $M\simeq10^{15}$gr is about $10^{-11} \text{Gyr}^{-1}$.  This calculation assures that probability of primordial black holes being trapped inside the earth is almost zero.  

We also calculated the energy released by a black hole collision with earth and compare it with the impact of asteroids on the earth. We have shown that the accretion process by the black holes is the dominant process for energy release. The amount of energy from this collision is comparable with a kilometer size asteroid where it happens four orders of magnitude more frequently than a black hole collision.  We conclude that the likelihood of the dangerous impact of primordial black holes with the earth is very low. 

\section*{Acknowledgments}
I would like to thank Shant Baghram for 
his useful comments. Also I would like to thank the anonymous referee for his/her helpful comments 
that improved this article. This research was supported by Sharif University of Technology’s Office of Vice 
President for Research under Grant No. G950214. 
 
{\bf Data Availability:}
 No new data were generated or analysed in support of this research.

 \label{lastpage}

\end{document}